\documentclass[preprint,preprintnumbers,amsmath,amssymb,nofootinbib]{revtex4}

\usepackage{mathrsfs}
\usepackage{etex}
\usepackage{amssymb,amsthm,amscd,amsbsy,array}
\usepackage{bm}% bold math
\usepackage{amsmath,dsfont}
\usepackage{soul} % underline, strikethrough, etc.
\usepackage{graphics,graphicx,xcolor}
\usepackage{tikz}
\usetikzlibrary{shapes,calc,arrows,arrows.meta,decorations.markings,angles,math}

\usepackage[colorlinks=true, pdfstartview=FitV, linkcolor=blue, citecolor=blue, urlcolor=blue]{hyperref} % hyperref

\newcommand{\red}{\textcolor{red}}

\newcommand{\blue}{\textcolor{blue}}

\newcommand{\gb}{\quad\colorbox{green}}

\newenvironment{redtext}{\color{red}}
{\ignorespacesafterend}
\newenvironment{bluetext}{\color{blue}}{\ignorespacesafterend}

\newenvironment{magentatext}{\color{magenta}}{\ignorespacesafterend}
\newenvironment{cyantext}{\color{cyan}}{\ignorespacesafterend}
\newenvironment{orangetext}{\color{orange}}
{\ignorespacesafterend}

\newcommand{\bmagenta}{\begin{magentatext}}
\newcommand{\emagenta}{\end{magentatext}}
\newcommand{\bcyan}{\begin{cyantext}}
\newcommand{\ecyan}{\end{cyantext}}
\newcommand{\bblue}{\begin{bluetext}}
\newcommand{\eblue}{\end{bluetext}}
\newcommand{\bred}{\begin{redtext}}
\newcommand{\ered}{\end{redtext}}
\newcommand{\borange}{\begin{orangetext}}
\newcommand{\eorange}{\end{orangetext}}

\numberwithin{equation}{section}

\let\ssection=\section
\renewcommand{\section}{\setcounter{equation}{0}\ssection}
\newcommand{\beq}{\begin{equation}}
\newcommand{\eeq}{\end{equation}}
\newcommand{\bec}{\begin{center}}
\newcommand{\ec}{\end{center}}

\newcommand{\KN}{{Kerr-Newman\;}}

\newcommand{\BH}{{Black Hole\,}}
\newcommand{\BHs}{{Black Holes\,}}

\newcommand{\bTheta}{\boldsymbol{\Theta}}

\newcommand{\bb}{{\mathbf{b}}}

\newcommand{\bx}{{\bm{x}}}

\newcommand{\bc}{{\mathbf{c}}}

\def\aand{{\quad\text{\and and}\quad}}
\def\where{{\quad\text{\small where}\quad}}

\def\ie{{\;\text{\small i.e.}\;}}

\renewcommand{\d}{\mathrm{d}}

\newcommand{\cE}{{\mathcal{E}}}

\newcommand{\cH}{{\mathcal{H}}}

\newcommand{\bp}{{\mathbf{p}}}

\newcommand{\SO}{\mathrm{SO}}

\newcommand{\mT}{{\mathscr{T}}}
\newcommand{\mH}{{\mathscr{H}}}

\newcommand{\cT}{\mathcal{T}}

\newcommand{\bv}{{\bf v}}

\def\bnabla{{\bm{\nabla}}}

\def\smallover\#1/\#2{\hbox{$\textstyle\frac{\#1}{\#2}$}} %

\def\bv{{\bm{v}}}
\def\bp{{\bm{p}}}

\def\parag{\hfil\break} %%%%% paragraph
\def\kikezd{\parag\underbar}

\def\bequ{\begin{enumerate}}
\def\eenu{\end{enumerate}}
\def\bitem{\begin{itemize}}
\def\eitem{\end{itemize}}

\def\beq{\begin{equation}}
\def\eeq{\end{equation}}
\def\beqa{\begin{eqnarray}}
\def\eeqa{\end{eqnarray}}

\def\barray{\left(\begin{array}}
\def\earray{\end{array}\right)}
\def\barraynb{\begin{array}}
\def\earraynb{\end{array}}

\def\IR{{\mathbb{R}}} %%%%% Reals
\def\IS{{\mathbb{S}}} % Round sphere
\def\?{{\,\gb{\fbox{\texttt{?}}\;}}\,}

\def\p{{\partial}}

\def \p{{\partial}}

\newcommand{\bE}{{\mathbf{E}}}

\def\benu{\begin{enumerate}}
\def\eenu{\end{enumerate}}
\def\bitem{\begin{itemize}}
\def\eitem{\end{itemize}}

\newcommand{\const}{\mathop{\rm const.}\nolimits}

\def\smallover#1/#2{\hbox{$\textstyle\frac{#1}{#2}$}} %
\def\smallcirc{{\raise 0.5pt \hbox{$\scriptstyle\circ$}}}
\def\cabove(#1){\stackrel{\smallcirc}{#1}}
\def\ccabove(#1){\,\stackrel{\smallcirc\smallcirc}{#1}\,}
\def\cccabove(#1){\stackrel{\,\smallcirc\smallcirc\smallcirc}{#1}\,}
\def\2{{\smallover1/2}}

\def\boxit#1{
\vbox{\hrule\hbox{\vrule\kern4pt
\vbox{\kern5pt#1\kern5pt}\kern4pt\vrule}\hrule}
} %%%%% boxit (Knuth)

\newcommand{\medbox}[1]{\fbox{%
\rule[-10pt]{0pt}{25pt}$\;\;\displaystyle{#1}\;\;$}%
}

\let\ssection=\section
\renewcommand{\section}
{\setcounter{equation}{0}\ssection}

\def\besub{\begin{subequations}}
\def\esub{\end{subequations}}
\begin{document}

%\preprint{\texttt{arXiv: 2207.06302v5 [gr-qc]}}

\title{Anyonic spin-Hall effect on the Black Hole horizon
}
\author{L.Marsot$^{1,2,3}$\footnote{marsot@cpt.univ-mrs.fr},
P.-M. Zhang$^{4}$\footnote{corresponding author. mailto:zhangpm5@mail.sysu.edu.cn},
and
P.~A. Horvathy$^{5}$\footnote{mailto:horvathy@lmpt.univ-tours.fr}
}

\affiliation{
${}^1$School of Science, Sun Yat-sen University, Shenzhen 518107, China
\\
${}^2$Aix Marseille Univ, Universit\'e de Toulon, CNRS, CPT, Marseille, France
\\
${}^3$Department of Physics, Babes-Bolyai University, Kogalniceanu Street, 400084 Cluj-Napoca, Romania
\\
${}^4$School of Physics and Astronomy, Sun Yat-sen University, Zhuhai 519082, China
\\
${}^5$ Institut Denis Poisson  CNRS/UMR 7013,  Universit\'e de Tours - Universit\'e d'Orl\'eans,\\ Tours 37200 (France)\\
%\yb{\fbox{MZH-Letter-II-6}}\\
}
\date{\today}

\begin{abstract}
\textit{
Using the fact that the horizon of black holes is a Carroll manifold, we show that an ``exotic photon'' i.e. a particle without mass and charge but with anyonic spin, magnetic moment and ``exotic'' charges associated with the 2-parameter central extension of the 2-dimensional Carroll group moves on the horizon of a Kerr-Newman Black Hole consistently with the Hall law.}
\bigskip

\noindent
Phys. Rev. D \textbf{106} (2022) no.12, L121503
doi:10.1103/PhysRevD.106.L121503
[arXiv:2207.06302 [gr-qc]].

\end{abstract}

\maketitle

\tableofcontents

%%%%%%%%%%%%%%%%%%%%%%%%%%%%%%%%%%%%%%%%%%%%%%%%%%%%%%

%%%%%%%%%%%%%%%%%%%%%%%%%%%%%%%%%%%
\section{Introduction
}\label{Intro}
%%%%%%%%%%%%%%%%%%%%%%%%%%%%%%%%%%%

The horizon of black holes is a genuine Laboratory to explore Gravitational Physics. Several interesting and non-intuitive effects, linked to key questions such as the Information Paradox \cite{Hawking:2015qqa,Hawking:2016msc}, are expected to take place on it. In this paper we add one more item to the list  by showing that by ``exotic photons'' (to be introduced below)  exhibit the \emph{Spin Hall effect} on the horizon.

The clue is  \emph{Carroll symmetry}.
The Carroll group, a ``degenerate cousin'' of the Galilei group (as put by L\'evy-Leblond \cite{Leblond}) is obtained by  contracting,  in the Poincar\'e group, the  velocity of light  to zero, instead of letting it go to infinity, as in the usual Galilean limit \cite{Leblond,SenGupta}. Alternatively, Carroll symmetry is found by restricting a Lorentzian space-time to a null hypersurface \cite{Carrollvs,DGH91,Carroll4GW,Bergshoeff14,Morand}.

Recent attention in the subject arose when Carroll symmetry was found to be relevant, for instance,
for physics on a black hole horizon \cite{Donnay16,MBHhorizon,DonnayM,Freidel22}~:  the celebrated BMS group is indeed conformal Carroll \cite{Bagchi,BMSCarr,ConfCarr}.
Interest in Carroll dynamics has long been limited, though, by that \emph{Carroll particles} (other than tachyons \cite{ConfCarr,deBoer:2021jej}) were  \emph{believed not to move} \cite{Leblond,SenGupta,Carrollvs,Bergshoeff14}.

As it will be explained elsewhere \cite{LongDraft}, ``no-motion''  is understood by studying deviations from null geodesics. The ``time'' coordinate of Carroll geometries is in fact a null coordinate from the ambient spacetime and  ``not moving'' means following the corresponding ambient null geodesic.
Another approach \cite{LongDraft,Bidussi} relates the immobility of quasiparticles called \emph{fractons} \cite{fractons} to  Carrollian boost invariance.

 The \emph{Anomalous Hall Effect} (AHE) observed in ferromagnetic crystals had been attributed to an anomalous current \cite{KarpLutt}. Later it was argued that spinning particles (including light) exhibit a  \emph{Spin-Hall effect} \cite{SPINHALL,LightHall}
for which a semiclassical explanation was proposed using a Berry phase--extended framework \cite{Chang:1995ebu,Chang:1995zz,AHEPLA,DHchiral}. The clue is the \emph{anomalous velocity relation},
\beq
\frac{d\bx}{dt} = \frac{\p\cE(\bp)}{\p\bp} - e\bE\times\bTheta\,,
\label{anoHvel}
\eeq
where $\cE(\bp)$ is the band energy, $\bE(\bx)$ the electric field, and $3$-vector $\bTheta(\bp)$ represents the Berry curvature. The anomalous velocity term here is clearly the mechanical counterpart of the anomalous current. Choosing $\bE$ in the $x-y$ plane and $\bTheta$ perpendicular to it, eqn. \eqref{anoHvel} reduces to the ``exotic'' Galilean model based on a \emph{two-parameter central extension} of the planar Galilei group \cite{LLGal,Galexo}. The physical relevance of central extensions was recognized by Bargmann \cite{Barg54}, followed by \cite{SSD}. See \cite{InzunzaPl} for another recent application. Extensions  hint at  deviations from null geodesics. For instance, the deviations of light from geodesic motions can be attributed to the coupling of photon spin to the gravitational field including gravitational waves \cite{Fermat,GosselinBM06,DuvalMS18,Frolov20,OanceaSHEL20,Carroll4GW,Yamamoto18,Andersson20,HarteOancea}.

It has recently been recognized that in $2+1$ dimensions the Carroll group admits  a 2-parameter central extension \cite{Azcarraga,Ancille} inducing an extended dynamics \cite{Marsot21}.

 The aim of this paper is to  study the Carrollian analog of the Galilean case, illustrated by motion on a specific Carroll geometry: the \KN \BH horizon.

%%%%%%%%%%%%%%%%%%%%%%%%%%%%%%%%%%
\section{Doubly extended Carroll particle}\label{DoubleCarroll}
%%%%%%%%%%%%%%%%%%%%%%%%%%%%%%%%%%

The two-parameter non trivial central extension of the planar Carroll algebra \cite{Azcarraga,Ancille} is given by,
\begin{align*}
& [J_3, K_i] = \epsilon_{ij} K_j\,, \quad
[K_i, K_j] = \epsilon_{ij} A_{exo}\,, \quad
[J_3, P_i] = \epsilon_{ij} P_j\,, \quad
[K_i, P_j] = \delta_{ij} P_0\,, \\
& [J_3, P_0] = 0\,, \quad
[K_i, P_0] = 0\,, \quad
[P_i, P_j] = \epsilon_{ij} A_{mag}\,, \quad
[P_i, P_0] = 0\,,
\end{align*}
where $J_3$ is the rotation, the $(K_i)$ are  boosts, the $(P_i)$ spatial translations, $(P_0)$ is time translation, and $A_{exo}$ and $A_{mag}$ are the ``exotic'' and ``magnetic'' extensions, respectively.
These extensions allow us to endow planar Carroll particles canonically  with two additional central charges we shall call, correspondingly, exotic and  magnetic and denote by $\kappa_{exo}$ and $\kappa_{mag}$, respectively \cite{Marsot21,LongDraft}.

%%%
 Then we show that massless uncharged particles (we call ``exotic Carrollian photon'') move on the horizon of a Kerr-Newman  Black Hole by \emph{following the Hall law}, providing us with a Carrollian version of the \emph{Spin-Hall effect} for anyons and contradicting the ``no-motion" statement \cite{Marsot21}.

Classical particle models associated with the transitive action of a given symmetry group are conveniently constructed using the Kirillov-Kostant-Souriau (KKS) orbit method \cite{KKS, SSD}. In Souriau's version that we follow here, the motions are determined by the ``Souriau two-form'' $\sigma$ which is closed and has constant rank. Then $\sigma=d\varpi$ (locally). The ``Cartan form'' $\varpi$ could be used for a variational calculus \cite{SSD,NCLandau}. Splitting the Souriau-form as $\sigma = \Omega -d\mH\wedge ds$ provides us, moreover, with a symplectic form $\Omega$ whose inverse  define in turn commutation relations \cite{NCLandau}.
Applied to the doubly-centrally-extended Carroll group, the construction yields for a free particle  with mass $m$ the Poisson brackets and Hamiltonian~\footnote{The commutation relations \eqref{CexoPB} have an overlap with those of the so-called ``Maxwell algebra'' \cite{Negro}. The respective \emph{Hamiltonians are substantially different}, though~: the doubly-extended Carroll and the ``Maxwell'' or ``enlarged" \cite{Enlarged} systems are fundamentally \emph{different} as they are built from different ingredients~: constant external electromagnetic fields for ``Maxwell'' and intrinsic central extension parameters for our doubly-extended Carroll, respectively.},
\beq
\{x_{i},x_{j}\}=\displaystyle\frac{\kappa_{exo}\;}{mm^*}\,\epsilon_{ij}\,,
\quad
\{x_{i},p_{j}\}=\displaystyle\frac{m\,}{m^*}\,\delta_{ij}\,,
\quad	
	\{p_{i},p_{j}\}=\displaystyle\frac{m\,}{m^*}\,\kappa_{mag}\,\epsilon_{ij}\,,
	\qquad
\mathscr{H}_0 \equiv 0\,,
\label{CexoPB}	
\eeq
where
\beq
m^*=
m \left(1 - \frac{\kappa_{exo}}{m^2}\kappa_{mag}\right)\,
\label{freeeffmass}
\eeq
is an effective mass, assumed not to vanish \cite{Marsot21,LongDraft}.

Let us underline that:
(i) the coordinates do not commute; (ii) the 2nd extension parameter $\kappa_{mag}$ behaves as an internal  magnetic field carried by the particle \footnote{In the Galilean theory \cite{Galexo} the mass is part of the moment map.}.

Note that the free Hamiltonian has \emph{no kinetic term}. The Hamiltonian equations of motion
 are thus trivially that a [doubly extended] \emph{Carroll particle with non-zero effective mass $m^*\neq0$ does not move}.

%%%%%%%%%
The free system \eqref{CexoPB} is by construction invariant w.r.t. the action of the doubly extended Carroll group
\beq
\bx \to
A\bx+\bc
\quad
s \to s - \bb\cdot A\bx +f \,,
\quad
\bp \to  A\bp +m\bb
\label{freeCarract}
\eeq
where $A\in\SO(2), \bc,\bb\in\IR^2$, $f\in\IR$, cf. \# (3.15) in \cite{Marsot21}. Here $s$ is Carrollian time.
All this follows from the structure of the Carroll group upon applying the KKS algorithm.

Coupling such a particle to an electromagnetic field modifies both the symplectic structure and the Hamiltonian \cite{SSD,Sternberg,NCLandau}. For simplicity, we restrict our attention at an uncharged doubly-extended Carroll particle \footnote{Charged particles are studied in \cite{LongDraft}.} with magnetic moment $\mu$ and anyonic spin $\chi$ in a static e.m. field $(B,\bE)$. The Poisson brackets are as in \eqref{CexoPB} but the modified  Hamiltonian $\mH=\mu \chi B$,
 yields, for $m\neq0$,
\beq
({x}^i)^{\prime} = \mu\chi\,\frac{\kappa_{exo}}{\kappa_{exo}\kappa_{mag} - m^2}\, \epsilon^{ij}  \p_jB\,,
\qquad
{p}_i^{\prime} = m^2\frac{\mu\chi \p_iB}{m^2-\kappa_{exo}\kappa_{mag}}\,
\label{IV.22BIS}
\eeq
where the prime means $d/ds$.
\goodbreak
These equations are both of the first order. The one for $\bx$ is fully decoupled and can be solved on its own, but the one for $\bp$ depends on the result for $\bx$.

Moreover, letting here $m\to 0$  \emph{$\kappa_{exo}$ drops out as long as it does not vanish}, leaving us with
\beq
\medbox{
({x}^i)^{\prime} = (\mu\chi)\,\epsilon^{ij}\frac{\p_jB}{\kappa_{mag}}
\aand
{p}_i^{\prime}=0\,.\\
}
\label{e0m0k}
\eeq
Note that the usual electromagnetic terms were switched off by choosing $e=0$, but the magnetic field plays a new role -- that of an electric potential.
 
Motion in a curved Carroll manifold was considered in \cite{Bergshoeff14,Marsot21},
however the gravitational field does not couple to the Carrollian equations of motion. An intuitive justification is that gravitational minimal coupling impacts, through the covariant derivative, the equation for the momentum, however not that for the position. However (as noticed  above) the momentum has no impact on the $\bx$-motion.

%%%%%%%%%%%%%%%%%%%%%%%%%%%%%%%%%%%%%%%%%%%
\section{Carroll structure of the Kerr-Newman horizon}
\label{KNhorizon}
%%%%%%%%%%%%%%%%%%%%%%%%%%%%%%%%%%%%%%%%%%%

A Kerr-Newman Black Hole characterized by
its mass $M$, angular momentum $J$, and charge $Q$   can be described by using the Eddington-like coordinates $(u,r,\vartheta, \phi)$ \cite{Newman65}. In these coordinates, the  metric
\besub
\begin{align}
\label{kn_g}
g = & - \frac{\Delta}{\Sigma} \left(du + \frac{\Sigma}{\Delta} dr - a \sin^2 \vartheta d\phi\right)^2 + \frac{\sin^2 \vartheta}{\Sigma} \left(a du - (r^2+a^2) d\phi\right)^2 + \Sigma d\vartheta^2 + \frac{\Sigma}{\Delta} dr^2
\\
&\Sigma =\; r^2 + a^2 \cos \vartheta,
\quad
\Delta = r^2 + a^2 + Q^2 - 2Mr \,,
\label{KNcoeff}
\end{align}
\esub
where $a = J/M$,
and its inverse are regular  on the  (outer) horizon $\cH$ of a Kerr-Newman black hole defined by $r = r_+=M+\sqrt{M^2-(a^2+Q^2)}=\const$ hypersurface defined by
$
\Delta = 0\,.
$
Note that the seemingly problematic $dr^2$ terms in \eqref{kn_g} containing $\Delta$ in their denominator cancel one another out.
Then we consider the $2+1$ dimensional structure \cite{DonnayM,MBHhorizon}
whose ingredients are the induced metric and a vector,
\besub
\begin{align}
\label{kn_g_h}
\widetilde{g} = &\;g|_{\Delta = 0} = \frac{\sin^2 \vartheta}{\Sigma} \left(a \, du - (r^2+a^2) d\phi\right)^2 + \Sigma d\vartheta^2\,,
\\[4pt]
\label{kn_xi}
\xi = & \;\partial_u + \Omega_H \partial_{\phi}
\where \Omega_H = \frac{a}{r^2+a^2}\,,
\end{align}
\esub
respectively. Here
$\Omega_H$ is the angular velocity of the horizon.
The restricted metric \eqref{kn_g_h} is singular as made manifest by the coordinate change
$ (\vartheta, \phi, u) \mapsto (\vartheta, \varphi = \phi  - \Omega_H u, u)\,,
$
which leads to the metric,
 %%%%%%%%%%%%%%
\begin{equation}
\widetilde{g} = \frac{(r^2+a^2) \sin^2 \vartheta}{\Sigma} d\varphi^2 + \Sigma d\vartheta^2 \,
\qquad \& \qquad
\xi = \partial_u\,,
\label{KNhmetric}
\end{equation}
The kernel is generated by the vector  $\xi$, $\widetilde{g}(\xi) = 0$.
%%%%%%%%%%%%%
Thus we have a degenerate metric and a vector field in its kernel, allowing us to conclude that \emph{the horizon $\cH$ of a Kerr-Newman black hole carries a Carroll structure} $(\IS^2 \times \IR, \widetilde{g}, {\xi})$  \cite{Carrollvs}.
The degenerate ``metric'' $\widetilde{g}$ carries the geometric information of the $\IS^2$ part of the black hole, while $\xi$ generates the $\IR$ part.

The horizon of a Kerr-Newman Black Hole carries a magnetic field
\beq
{B} =
\left(2 a Q r_+(r_+^2+a^2)\right) \,\frac{\cos\vartheta}{\left(r_+^2 + a^2 \cos^2\vartheta\right)^3}\,,
\label{Bonhor}
\eeq
while in comoving coordinates the electric field vanishes \cite{LongDraft}.

%%%%%%%%%%%%%%%%%%%%%%%%%%%%%%%%%%%%%%%%%%
\section{Motion on the Kerr-Newman horizon}\label{hormot}
%%%%%%%%%%%%%%%%%%%%%%%%%%%%%%%%%%%%%%%%%%

A \emph{massive particle associated with the unextended Carroll group can  stay fixed, but can not move} \cite{Carrollvs,ConfCarr}. However the horizon is a $2+1$ dimensional Carroll manifold, therefore the particle may have an extended dynamics associated with the double central extension with parameters
 $\kappa_{exo}$ and $\kappa_{mag}$ \cite{Marsot21,Azcarraga,Ancille,LongDraft}.

 Now we show that the extended dynamics \emph{can} lead to motion, namely on the \BH horizon.
  Remember first that geodesics on the horizon are necessarily massless \cite{Carrollvs} \footnote{Such a photon trajectory could be created by turning on a lamp and then throwing it into a stationary black hole in such a way that when the lamp crosses the horizon, the photons be emitted in the direction of the horizon's null generator.}.
 An ``exotic photon'' i.e. one with \emph{no mass and charge}, $m=0$ and $e=0$, but with nonvanishing magnetic moment $\mu$, anyonic spin $\chi$ and double central extension can be coupled to the electromagnetic field through a spin-field term
$\mathscr{H}=\mu \chi\, B$ where $B$ is the magnetic field \eqref{Bonhor} on the horizon. Then the equations of motion \eqref{e0m0k}  describe an Anomalous
 Spin-Hall Effect with $\bnabla B$ behaving as an \emph{effective electric field}, ${\mu \chi}$ as an effective electric charge, and $\kappa_{mag}$ as an effective magnetic field.

Coupling to the gravitational field amounts to  replacing the derivative on $p_i$  by a covariant derivative  \cite{Marsot21}. However, this does not change the velocity equation which is indeed the only relevant one for the poor Carrollian dynamics: the momentum equation remains decoupled.

Having a non-zero gradient for the magnetic field \eqref{Bonhor} requires non-zero electric charge $Q$ and angular momentum $J$ (since $a=J/M$).

Using (comoving) angular coordinates $(\vartheta, \varphi, u)$ we see that the electric field induced on the horizon vanishes. The radial component would survive, but disappears in the 2+1 restriction.
The gradient of $B$ is tangent to the
longitudinal great circles $\varphi=\const$
By \eqref{e0m0k}, the motion is governed by,
\begin{equation}
\medbox{
({x}^\vartheta)^{\prime} = 0 \, ,
\quad
\,({x}^\varphi)^{\prime} = \Big(2 a Q\, r_+\,(r_+^2+a^2)\,
\frac{{\mu \chi}}{\kappa_{mag}}\Big)\,
 \frac{(r_+^2 - 5 a^2 \cos^2 \vartheta)}{(r_+^2 + a^2 \cos^2 \vartheta)^4} \sin \vartheta \, .
}
\label{kn_eom_h}
\end{equation}
%%%%%%%%%%
%%%%%%%%%%
\begin{figure}[!ht]
\includegraphics[scale=0.4]{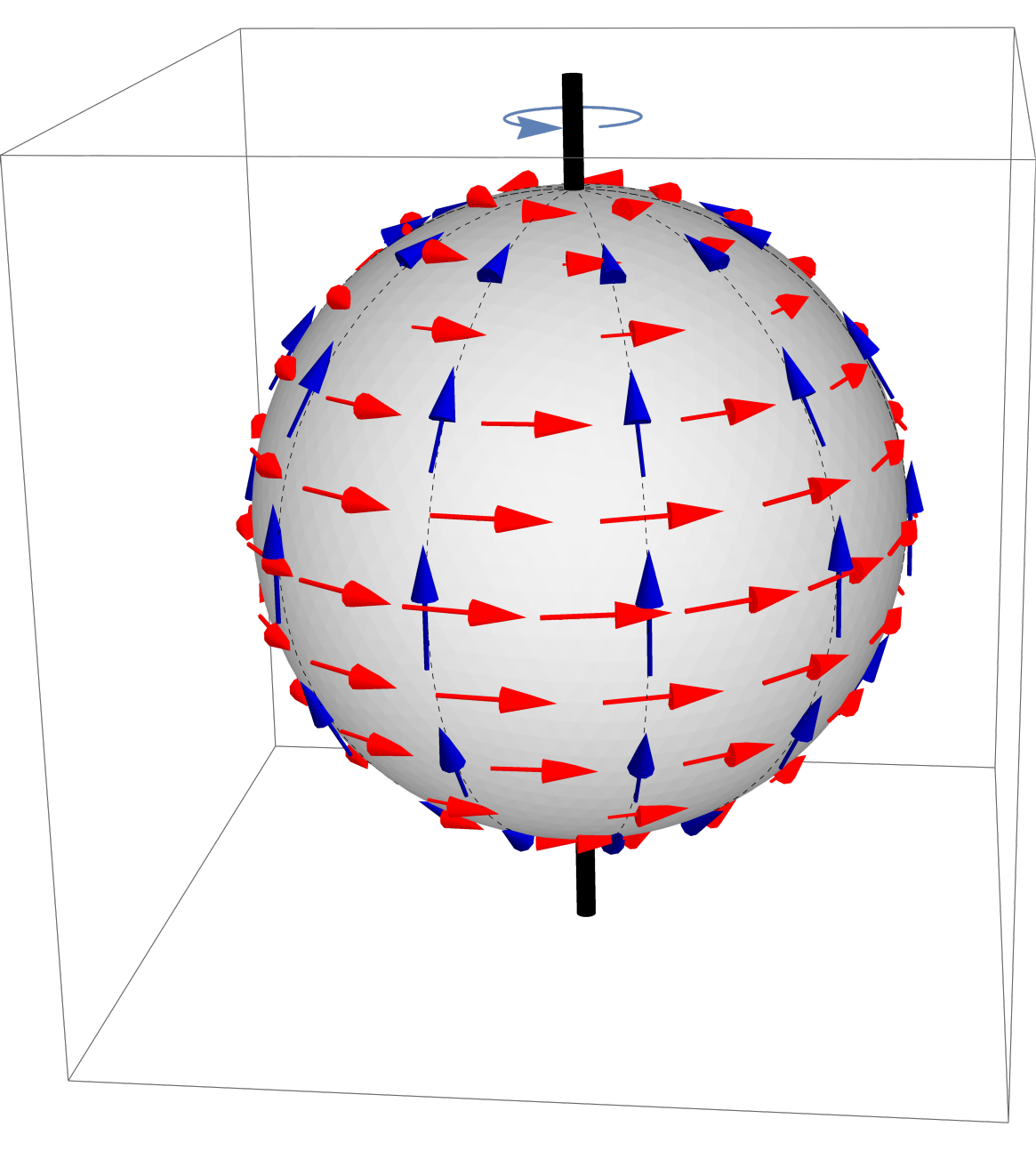}\vskip-3mm
\caption{\it {\small On the horizon of a Kerr-Newman Black Hole the \red{\bf velocity field}  \eqref{kn_eom_h}
is perpendicular to the axis of rotation and obeys  the Anomalous Hall law with an \blue{{\bf effective electric  field} $\bE^*=\bnabla B$} which is  tangent to the longitudinal great circles with $\mu\chi$ playing the role of an effective electric charge.
The  arrows indicate the directions and norms.
 }
\label{f:bh_drift}
}
\end{figure}
%%%%%%%%%%%
Thus our ``exotic photon'' performs  azimuthal circular motion  with  $\vartheta=\const$, parameterized by $\varphi$. Consistently with the Hall behavior, the motion is perpendicular to the (longitudinal) effective electric field $\bnabla B$ (which vanishes at the poles and takes its maximum on the equator).
The direction of the rotation is correlated with the angular momentum $J$ and the charge $Q$ which should not vanish --- and this is precisely the reason that we consider \KN \BHs.
The angular velocity  goes smoothly to zero as we approach the poles and depends on the radius of the horizon roughly as $r_+^{-3}$, implying that the rotation would be more important for smaller black holes.

The rotation we have just found, although reminiscent of the frame-dragging by a rotating black hole,  is however \emph{unrelated} to it: frame-dragging is hidden in the coordinates, which are comoving with the horizon.

%%%%%%%%%%%%%%%%%%%%%%%%%%%%%%%%%%%%%%%%%%%%%
\section{Carroll symmetry on the horizon}\label{horCarr}
%%%%%%%%%%%%%%%%%%%%%%%%%%%%%%%%%%%%%%%%%%%%%

We conclude our Letter with a short survey of the {symmetries}, conveniently studied by looking at the Cartan $1$-form  \cite{SSD}
  $\varpi$ defined by $\sigma=\d\varpi$ as mentioned in sec.\ref{DoubleCarroll}. Switching to $\bv = \bp/m$ before letting $m\to0$ yields
\begin{equation}
\varpi = \frac{\kappa_{exo}}{2} \epsilon_{ij} v^i dv^j + \frac{\kappa_{mag}}{2} \epsilon_{ij} x^i dx^j + \mu \chi B ds\,.
\end{equation}
 In presymplectic terms,
 the Noether theorem says: \textit{a vector field $X$ is a symmetry of the dynamics, if $L_X \varpi$ vanishes up to a total derivative, $L_X \varpi= df$. Then
Cartan's formula implies that
\beq
Q_X=
i_X \varpi- f
\label{SimpNoether}
\eeq
is conserved} \cite{SSD}.

 The isometry group of the \KN horizon  $\cH$ is  $\SO(2) \ltimes \cT$, generated by the vectorfields
\begin{equation}
X = \partial_{\widetilde{\varphi}} +
\mT(\vartheta, \widetilde{\varphi}) \partial_s\,,
\end{equation}
where the ``supertranslation'' $\mT$ is an arbitrary function of the coordinates ($\vartheta, \widetilde{\varphi}$) on the horizon.
Thus~:
%%%%%
\begin{itemize}
\item
Translations of the black hole horizon generated by $\p_\varphi$ change $\varpi$ by a surface term, $L_{\p_\varphi} \varpi = df$ with $f = - (\kappa_{mag}/{2}) \vartheta$. Thus \eqref{SimpNoether} yields the conserved quantity,
\beq
p_\varphi = \kappa_{mag} \vartheta\,.
\label{pfi}
\eeq
This unusual expression is consistent with  (3.18c) in \cite{Marsot21}.
\item
Now look at the zeroth order expansion of a supertranslation, \ie a  (Carrollian) time translation $X = \partial_s$. We readily have $L_X \varpi = 0$, and so
\begin{equation}
i_X \varpi = \mu \chi B \equiv \mH\,,
\label{consH}
\end{equation}
identified with the Carroll Hamiltonian is conserved.

\item
For a general supertranslation $X = \mT(\vartheta, \varphi) \partial_s$ we have, instead,
\begin{align}
L_X \varpi =& (\mu\chi) B \partial_i \mT dx^i,
\label{Tthetaphi}
\end{align}
which is \emph{not} a total derivative in general due to  $d(L_X \varpi) \propto dB\wedge d\mT\neq0$ unless $\mT=\mT(\vartheta)$ --- for which \eqref{SimpNoether} then yields a conserved quantity.
 If the supertranslation is, for example, induced by  the magnetic field,
$
\mT =\mT(B)$ e.g. for
$\mT_n \propto B^n$
for some positive integer $n$, then $L_X \varpi=n\, d \mH^{n+1}$
 is a total derivative, providing us with an infinite tower of conserved quantities
$
Q_n
= \mH^{n+1},
\,
$
--- which are however mere powers of the Hamiltonian in \eqref{consH}.
\item
%%%%

It is instructive to study Carroll boosts in \eqref{freeCarract},
\beq
\bx \to \bx \qquad
s \to s - \bb\cdot\bx\,, \qquad \bb\in\IR^2\,,
\label{Cboost}
\eeq
characteristic for the Carroll symmetry. They  belong  to the isometry ``bottom'' of  BMS supertranslations \cite{BMSCarr}.
``Horizontal'' boosts  along $\p_\varphi$ are broken by the magnetic field,
however for ``vertical'' boosts,
$\mT = -b_{\vartheta}\vartheta$,
 \eqref{SimpNoether} provides us with \footnote{For $B=\const$ we would get $Q = 0$ consistently with \cite{ConfCarr}.}
\beq
Q= (\mu\chi)\left[-B\vartheta + \int\! B d\vartheta\right]\,.
\label{vboostCQ}
\eeq

\end{itemize}

So far we we proceeded as follows: first we solved the equations of motion and then  checked that the associated Noetherian quantities are indeed conserved along the trajectories. The conservation of $p_{\vartheta}$ in \eqref{pfi}, of $\mH$ in \eqref{consH} or of even of the weird boost-momentum in \eqref{vboostCQ} is indeed manifest from that
\beq
\vartheta=\const
\label{thetaconst}
\eeq
along the trajectories as we had found earlier.

However conservation laws are often used  conversely, i.e., to derive the motions. Can we proceed in the reversed direction ?  Remarkably, the answer is YES:  their explicit forms manifestly \emph{require} \eqref{thetaconst} for being conserved~!

\goodbreak
%%%%%%%%%%%%%%%%%%%
\section{Conclusion}
%%%%%%%%%%%%%%%%%%%

The absence of the kinetic term in their Hamiltonian implies that Carroll particles have a purely anomalous velocity relation. Position and momenta are partically decoupled and we end up with  first-order equations. The motion of an ``exotic photon'', \eqref{e0m0k},
 is poor but \emph{not entirely trivial}: it exhibits the
 \emph{Anyonic Spin-Hall effect} \cite{LightHall}.

The horizon of a Kerr-Newman Black Hole realizes these conditions: its  magnetic field  $B$  \eqref{Bonhor}  induces  anomalous Hall motion for our ``exotic photon".
Masslessness is mandatory for getting non-trivial dynamics \cite{LongDraft}.

Particles of the type of our ``exotic photons" might actually play a r\^ole in condensed matter physics as quasiparticles \cite{Zoo}. Here we took them chargeless for simplicity, however they could in principle carry also an electric charge \cite{LongDraft}.

The double central extension of the Carroll group is a mathematical fact \cite{Azcarraga,Ancille}. But is it
 a physical reality ?
With no  experimental data at hand, we just recall Dirac about his magnetic monopole \cite{Dirac31}: \begin{quote}\textit{\narrower
``This new development \dots is merely a generalisation of the possibilities \dots Under these circumstances one would be surprised if Nature had made no use of it.''
}
\end{quote}
We note also that our gravitational ideas could in principle be tested in Laboratory by the remarkable analog of a \KN \BH which could be created in condensed matter  \cite{opticalKN,Soft}.

\kikezd{Note added}.
After the revised version of our paper has been resubmitted we were informed by T. R. Perche  \cite{Kubiznak} that they are also considering similar problems and arrived, by following similar methods, to similar but slightly different conclusions.

\begin{acknowledgments}\vskip-4mm
We are grateful to J. Balog, G. Gibbons, M. Chernodub, M. Stone, P. Chru\'sciel, M.~Plyushchay and  Z. Silagadze for advice. LM is thankful to the Physics Department of Babe\c{s}-Bolyai University for hospitality.
PMZ was partially supported by the National Natural Science Foundation of China (Grant No. 11975320).
\end{acknowledgments}
\goodbreak

%%%%%%%%%%%%%%%%%%
%%%%%%%%%%%%%%%%%%
\end{document}